\documentclass{elsart}
\usepackage{amssymb,epsfig}
\usepackage{graphicx}

\newcommand{\beq}[1]{
\begin{equation}\label{#1}}
\newcommand{\eeq}{\end{equation}}
\newcommand{\bea}[1]{
\begin{eqnarray}\label{#1}}
\newcommand{\eea}{\end{eqnarray}}
\newcommand\re[1]{(\ref{#1})}
\newcommand{\beqar}[1]{\begin{eqnarray}\label{#1}}
\newcommand{\eeqar}{\end{eqnarray}}
\usepackage{psfrag}

\psfrag{kt}{{\footnotesize $k_\perp$}}
\psfrag{=}{{\footnotesize $=$}}
\psfrag{+}{{\footnotesize $+$}}
\psfrag{l}{{\footnotesize $l$}}
\psfrag{p1z'}{{\footnotesize $p_1z'$}}
\psfrag{q1}{{\footnotesize $q_1$}}
\psfrag{p1}{{\footnotesize $p_1$}}
\psfrag{-p1bz'}{{\footnotesize $-p_1{\bar z}'$}}
\psfrag{-q2}{{\footnotesize $-q_2$}}
\psfrag{x1p2}{{\footnotesize $x_1p_2$}}
\psfrag{x2p2}{{\footnotesize $x_2p_2$}}
\psfrag{p2}{{\footnotesize $p_2$}}
\psfrag{p2'}{{\footnotesize $p_2'$}}
\psfrag{kpx1p2}{{\footnotesize $k_\perp, x_1$}}
\psfrag{kpx2p2}{{\footnotesize $k_\perp, x_2$}}
\psfrag{pi}{{\footnotesize $\pi$}}
\psfrag{a}{{\footnotesize $(a)$}}
\psfrag{b}{{\footnotesize $(b)$}}
\psfrag{c}{{\footnotesize $(c)$}}
\psfrag{d}{{\footnotesize $(d)$}}
\psfrag{e}{{\footnotesize $(e)$}}
\psfrag{f}{{\footnotesize $(f)$}}
\psfrag{g}{{\footnotesize $(g)$}}
\psfrag{Fipi}{{\footnotesize $\phi_\pi(z')$}}
\psfrag{Fchi}{{\footnotesize $F_\xi(x_1)$}}

\psfrag{I2}{{$|{ I}|^2$}}

\psfrag{z}{{$z$}}
\psfrag{TH}{\footnotesize $T_H(z',x_1)$}

\begin{document}
\begin{frontmatter}
\title{OCD factorization for the pion diffractive dissociation into two
jets
\thanksref{talk}
}
\author{L.~Szymanowski
}
\thanks[talk]{This talk is based on results obtained in collaboration
with V.M.~Braun, D.Yu.~Ivanov and A. Sch\"afer.}

\address{So{\l}tan Institute for Nuclear Studies, 
Ho\.za 69, Warsaw, Poland}
\address{Institut f\"ur Theoretische Physik, Universit\"at Leipzig, Germany}
\ead{lechszym@fuw.edu.pl}
\begin{abstract}
We present main ideas and obtained results  of our recent calculations  of  Coulomb and QCD contribution to 
the pion diffractive dissociation into two jets.
\end{abstract}
\begin{keyword}
Perturbative QCD

\PACS 12.38.Bx
\end{keyword}
\end{frontmatter}

\section{Introduction}

This talk  is devoted to our studies of
the pion diffractive dissociation into two jets, being the  subjects of Refs.
 \cite{IS},
\cite{BISS}.
In this process
a highly energetic pion interacts
with a nucleus $A$ and produces
two jets (di-jets) which in the lowest approximation are formed from the
quark-antiquark ($q\;\bar q$) pair
\begin{equation}
\label{process}
\pi\;A\;\rightarrow\;q\;\bar q\;A\;\;.
\end{equation} 
Primary motivation for studies of this process  goes back to Ref.
\cite{BBGG81}, in which
 it has been conjectured 
that pion diffraction dissociation on a heavy nucleus $\pi A\to XA$
is sensitive to small transverse size configurations of pion constituents.
It was later argued \cite{FMS93}, \cite{NSS00}, \cite{FMS00} that selecting a specific hadronic
final state that consists of a pair of (quark)  jets with large transverse
momentum 
one can obtain important
insight into the pion
structure as it turns out that the longitudinal momentum fraction
distribution of the jets follows that of the pion valence parton
constituents. A measurement of hard dijet coherent production on nuclei
presents, therefore, a possibility of a direct measurement
of the pion distribution amplitude and provides
striking evidence \cite{exp00} that this distribution is close to
its asymptotic form.

From the theoretical point of view, the principal question is whether
the relevant transverse size of the pion $r_\perp$
(alias the scale of the pion distribution amplitude $\mu = 1/r_{\perp}$)
is 
of the order of the
transverse momenta of the jets $\mu \sim q_\perp$. In 
this case one
can write the scattering amplitude as a convolution of the nonperturbative
soft pion distribution amplitude and the skewed gluon distribution in the
target with  the hard scattering amplitude.

The cross section for this process is largest  when
 the momentum
transfer to the target is the smallest one, which means that the
large
transverse momenta of the quark jet ($q_{1\perp}$) and  of the antiquark
one ($q_{2\perp}$) have to balance each other 
$q_{1\perp}\simeq -q_{2\perp}$.

The main contribution to the diffractive process (\ref{process})
is due to  Pomeron exchange. However,
 this process can also occur as result of the electromagnetic
interactions between pion and  target nucleus (Coulomb exchange).
The strenght of the electromagnetic coupling
is
 $\alpha\,Z$ ($\alpha$ is the electromagnetic fine coupling constant).
For heavy nuclei $\alpha\;Z$ is not small and therefore one can ask about
the size
of the
Coulomb contribution to (\ref{process}). Since  derivation of the Coulomb 
contribution is 
technically simpler let us begin from this case.

\section{Coulomb dissociation of a pion into two jets }

The gauge invariant set of 
diagrams describing the Coulomb contribution to the process
(\ref{process})
 is shown in Fig. 1.
%
\begin{figure}[h]
\begin{center}
\includegraphics*[width=7cm]{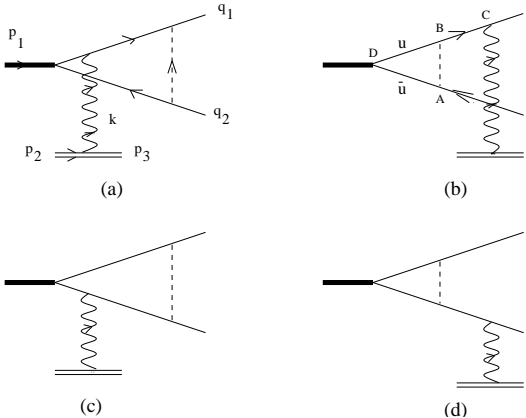}
\end{center}
\caption{\label{pifig1}
Coulomb contribution  to pion dissociation into two jets.
}
\end{figure}
%
The wavy line denotes  photon exchanged in the t-channel.
The large
transverse momentum of di-jets results from the hard gluon exchange
denoted by  the dashed line.
It supplies a hard scale to the process so we neglect the pion
mass.
The incoming pion has momentum  $p_1,\; p_1^2=0$ and its fraction
carried by quark (antiquark) is denoted by $z$ ($\bar z \equiv 1-z$).
The nucleus mass is $M$, its momenta in the final
and in
the initial states are  $p_2$ and $p_3$, respectively.
The large cms energy squared of the process is 
$s=(p_1+p_2)^2= \bar s +M^2$, and the small momentum transfer squared
equals to the photon virtuality $k^2$ , i.e.
$t=k^2=(p_2-p_3)^2$.
The Sudakov decomposition of the photon momentum $k$ 
\begin{equation}
\label{Sudk}
k= \alpha p_1 + \beta p_2^\prime + k_\perp\,,\;\;\;p_2'=p_2-\frac{M^2}{\bar
s}\,,\;\;\;p_2'^2=0\,,
\end{equation}
is fully determined by the kinematics of the process and can be written in terms
of the jets variables as
\begin{equation}
\label{albe}
\beta =\frac{M^2_{2j}+(q_{1\perp}+q_{2\perp})^2}{\bar s} \,,\;
\alpha=-\frac{(q_{1\perp}+q_{2\perp})^2+\frac{M^2\beta}{\bar s}}{\bar s}
\,,\;
k_\perp =q_{1\perp}+q_{2\perp}\,,
\end{equation}
where $M^2_{2j}$ is  the invariant mass of the di-jets.
The momentum transfer  $t$
 equals $t=k^2 \approx -(k_\perp^2+k_{min}^2)$, and according to
Eq.(\ref{albe}) its minimal value  is
\begin{equation}
\label{kmin}
k_{min}^2= \frac{M^2M_{2j}^4}{\bar s^2}\,, \;\;\;\;\;\mbox{where}
\;\;\;\;\;\;M^2_{2j}=\frac{q^2_{1\perp}}{z\bar z}\;.
\end{equation}

We calculate the leading asymptotics of the scattering amplitude
$M$
in powers of $1/q^2_{1\perp} $ (at the leading twist level).
It is given  as the convolution of the hard scattering
amplitude $T_H(u,\mu_F^2)$ with the pion light-cone
distribution amplitude $\phi_\pi (u,\mu^2_F)$
\begin{equation}
\label{conv}
M=\int\limits_0^1\;du\,\phi_\pi (u,\mu^2_F)\;T_H(u,\mu_F^2)\,,
\end{equation}
where the hard scattering amplitude $T_H(u,\mu_F^2)$ describes
the production of a free $q \bar q$ pair (the di-jets) in collision of
the $t$-channel photon with  $q$ and $\bar q$
(the later having momenta $u p_1$ and $\bar u p_1$, respectively),
 collinear to the pion momentum
$p_1$.
The factorization scale $\mu_F$ is of the
order of the di-jets transverse momentum $q_{1\perp}$. 

Eq.(\ref{conv})
describes  the factorization procedure in QCD  which disentangles the
contributions to $M$ coming from  large and small distances.
The soft part  is described
by the pion  distribution amplitude \cite{BL}, \cite{CZ}
\begin{equation}
\label{soft}
\langle 0|\bar d(x) \gamma_\mu
\gamma_5 u(-x)|\pi^+(p)\rangle_{x^2\to 0}=ip_\mu f_\pi
\int\limits^1_0 du e^{i(2u-1)(xp)} \phi_\pi (u) \ .
\end{equation}
The constant $f_\pi=131 MeV$ is known experimentally from $\pi\to \mu \nu$
decay.

The hard part, i.e. the amplitude $T_H(u)$ in Eq.(\ref{conv}),
 is calculable in the
perturbative QCD, it is given by four tree diagrams shown in Fig. 1.

The factorization procedure described by Eq. (\ref{conv}) is similar
to that  used for
 the hard exclusive processes \cite{BL}, \cite{CZ}.
Its validity for the quasi-exclusive process (\ref{process}) can be
justified as
follows.
 Let us consider, for example, the diagram  Fig.1(b).
The large transverse momentum of quark jets flows along the lines
 $A\; -\; B\;-\;C$. Therefore their virtualities are much larger than those
of  the other quark lines $D\;-\; A$ and $D\;-\; B$.
Thus, at  leading twist,
the quark lines $D\;-\; A$ and $D\;-\; B$ have to  be
considered as being on
mass shell and this part of the diagram
can be factorized out of the hard part
given by the
highly virtual quark and gluon propagators.

The result
for the scattering
amplitude corresponding to the diagrams in Fig.~1
is given by the formula (for details of derivation see Ref. \cite{IS})
\begin{eqnarray}
\label{fullamp}
&& M^{\pi^+ +A\rightarrow 2j + A} = -i
\delta_{i j}\,f_\pi\,\frac{2^6}{3^2}\,\pi^2\,\alpha\, Z\,F_{QED}(k^2)\,
\alpha_s(q_{1\perp})\,
\frac{\bar s}{(k_\perp^2 + k_{min}^2)\,(q_{1\perp}^2)^2} \nonumber \\
&& (e_u\,{\bar z} + e_d
\,z)\,
{\bar u}(q_1)\,\gamma^5\,
\left[ z\, {\hat k}_\perp \, {\hat q}_{1\perp} +
{\bar z}\, {\hat q}_{1\perp}
{\hat k}_\perp \right] \frac{{\hat p}_2'}{\bar s} v(q_2)
\int\limits_0^1\,du\,\frac{\phi_\pi(u)}{u}\;\;.
\end{eqnarray}
where $\alpha_s=\frac{g^2}{4\pi}$, and we used the symmetry property
$ \phi_\pi(u)=\phi_\pi(\bar u)$. The nucleus was here treated as a scalar
particle with the elctromagnetic form-factor $F_{QED}(k^2)$.

The conclusions which one can draw from Eq.(\ref{fullamp}) are the
following.
Since the behaviour of the pion distribution
amplitude at the end-points is known \cite{BL}, \cite{CZ}:
$\phi_\pi(u)\sim u$ for $u\to 0$, and $\phi_\pi(u)\sim \bar u$ for
$\bar u\to 0$, therefore
  the integral over the momentum fraction
fraction $u$ is well defined what
confirms that factorization holds
for the process  we discuss.
Let us emphasize that the integral
of $\phi_\pi(u)$ over $u$
in Eq. (\ref{fullamp}) generates
 only an overall factor.
Therefore the dependence of the amplitude $M$ on $z$ is universal, i.e.
 it doesn't depend on the shape of
the pion distribution amplitude.
Moreover, let us  note that the
amplitude (\ref{fullamp}) vanishes for $z=e_u/(e_u - e_d)=2/3$.
Due to the opposite signs of the electric charges of the pion constituents,
 $e_u=2e/3, e_d=-e/3$, the contribution of the
diagrams (a) and (b) in Fig. 1
cancels the one of diagrams (c) and (d).
 
The result (\ref{fullamp}) differs essentially
from
the corresponding formula (78) of Ref. \cite{FMS00}.
We predict the universal $z$ dependence of
the scattering amplitude $M$, independent of the shape of
$\phi_\pi(u)$. Contrary to that, in 
Ref. \cite{FMS00}
the amplitude is proportional to the lowest pion Fock state
wave function, $M \sim \Psi_\pi(z, q_{1\perp})$ (in our notation).
This disagreement between  two results is in our opinion related
 to unproper treatment of the
hard gluon exchange in Ref. \cite{FMS00}. Within the QCD factorization
approach,
which we followed in this study, the hard gluon exchange has to be
treated as a part of
the hard scattering block $T_H(u,\mu^2_F)$. In Ref. \cite{FMS00}
the hard gluon exchange is considered as
the high transverse momentum tail  of the wave function
$\Psi_\pi(z, q_{1\perp})$.

Our  estimates of the magnitude of the Culomb effect show 
that it is very small in comparison to the QCD contribution 
(for details
see Ref. \cite{IS}).  
 But 
we cannot directly compare our predictions 
with E791 data since their absolute normalization unfortunately is
not reported.

\section{QCD contribution to the pion dissociation into two jets}

\begin{figure}[t]
\begin{center}
\includegraphics*[width=7cm]{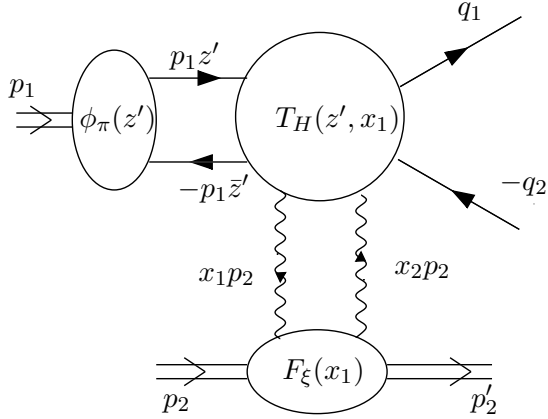}
\end{center}
\caption[]{\small
Kinematics of the coherent hard dijet production $\pi\to 2\, {\rm jets}$.
The hard scattering amplitude $T_H$ contains at least one hard gluon
exchange.
 }
\label{fig:1}
\end{figure}

The kinematics of the process is shown in Fig.~\ref{fig:1}.
We restrict ourselves to scattering from a single nucleon
having
initial momentum  $p_2$ and the final one $p_2^\prime$.
%
%
Fig.~\ref{fig:1} expresses also graphically the factorization formula
for the amplitude
of hard dijet production in QCD
\beq{factor}      
{
M}_{\pi \to 2\, {\rm jets}}
=\int\limits^1_0 dz^\prime \int\limits^1_0 dx_1 \,\phi_\pi
(z^\prime,\mu_F^2)\,T_H(z^\prime , x_1, \mu_F^2)\,F^g_\zeta(x_1, \mu_F^2)\,.
\eeq
in which $\phi_\pi (z^\prime,\mu_F^2)$ is again as in the Coulomb case 
(\ref{conv}) 
the pion distribution amplitude,
and $F^g_\zeta(x_1, \mu_F^2)$ is the non-forward (skewed) gluon distribution
\cite{M94,Rad96,Ji97} 
in the target nucleon or nucleus.
(The asymmetry parameter $\zeta$ is fixed by the process kinematics.)
$T_H(z^\prime , x_1, \mu_F^2)$ is the hard scattering amplitude
and $\mu_F$ is the (collinear) factorization scale.
 
Since at high energies the scattering amplitudes corresponding to
Pomeron exchange are dominated by their imaginary parts
we calculated only cut diagrams. 
They are
built of tree-level on-shell scattering amplitudes and their form is
strongly constrained by gauge invariance.
%
\begin{figure}[t]
\begin{center}
\includegraphics*[width=10cm]{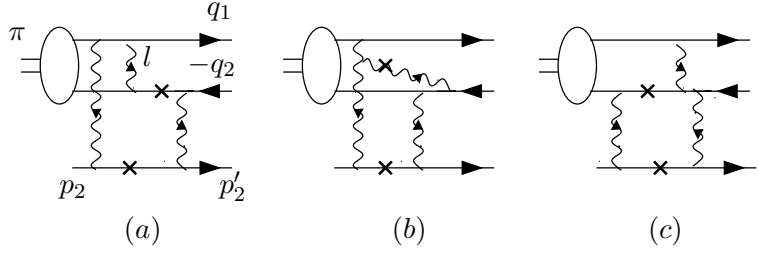}
\end{center}
\caption[]{\small
Cut diagrams (examples) for the imaginary part of the amplitude
$\pi q\to (\bar q q)q$. The cut quark (gluon) propagators are indicated
by crosses.
 }
\label{fig:2}
\end{figure}
%
The existing cut diagrams can be  grouped into the  four
gauge-invariant contributions shown in Fig.~\ref{fig:3}a--d,
which differ by the position of the hard gluon that provides the large
momentum transfer to the jets.
For example, in  Figs.~\ref{fig:3}a and 4b it is
assumed that the hard gluon exchange appears to the left of the cut and to
the right, respectively;
typical diagrams are  shown
in Figs.~\ref{fig:2}a and 3c. 
The two remaining contributions in Fig.~\ref{fig:3}c
and Fig.~\ref{fig:3}d take into account the possibility of
real gluon emission in the intermediate state. The filled circles
stand for the effective vertices describing the gluon
radiation.
%
\begin{figure}[htbp]
\begin{center}
\includegraphics*[width=11cm]{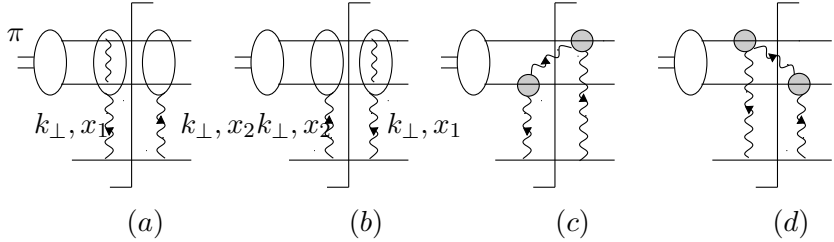}
\end{center}
\caption[]{\small
The decomposition of the imaginary part of the amplitude $\pi q\to (\bar q
q)q$
into four gauge-invariant contributions.
 }
\label{fig:3}
\end{figure}
%
The calculations are performed in  Feynman gauge and for details of them I
refer to Ref. \cite{BISS}. 
%

The final result for the imaginary part of the amplitude for
dijet production from a nucleon reads
\beq{ImM}
 {\rm Im}\,{
 M} =
  -i\,s\,f_\pi\,\alpha_s^2\,\frac{4\,\pi^3}{N_c^2\,q_{\perp}^4}\,{\bar
u}(q_1)\gamma_5 \frac{{\!\not\! p}_2}{s}v(q_2)\,{
 I}\, \delta_{i\,j}\,
\eeq
\bea{I}
&&{I} =
 \int\limits_0^1\!dz'\, \phi_\pi(z', \mu^2)
\left\{
%
\left(\frac{z\bar z}{z'\bar z'}+1  \right)
\left[C_F\left(\frac{z\bar z}{z'\bar 
z'}+1\right)+\frac{1}{2N_c}\left(\frac{z}{z'}
+\frac{\bar z}{\bar z'} \right)\right] \right. \nonumber \\
&&\left. \times\left[\frac{\Theta(z'-z)}{(z'-z)}\,F_\zeta\left(\frac{\zeta\,z'\bar
z}{z'-z},\mu^2
\right) + \frac{\Theta(z-z')}{(z-z')}\,F_\zeta\left(\frac{\zeta\,\bar z'
z}{z-z'},\mu^2  \right)    \right]\right. \nonumber \\  
&&\left. +\left[C_F\left(\frac{z\bar z}{z'\bar z'}-1\right)
\left(\frac{\bar z}{z'}+\frac{z}{\bar z'}\right)
+\frac{1}{2N_c\,z' \bar z'}\left(\frac{z\bar z}{z'\bar z'}+1\right)
\right]\,F_\zeta(\zeta,\mu^2) 
\right\}
\eea
Let us discuss conclussions which follow from  this result.
We begin by noting  that similarly as in the Coulomb contribution, the scattering
amlitude $M$ is not directly proportional to the pion distribution amplitude
$\phi_\pi(z, \mu^2)$ but it is a rather complicated convolution of this
quantity and the corresponding coefficient function.
The singularity at $z'=z$ of the integrand in \re{I} 
is present in the contributions in
Fig.~\ref{fig:3}c,d which include real gluon emission in the intermediate
state. The logarithmic integral $\int dz'/|z-z'| \sim \ln s$ is nothing but
the usual energy logarithm that accompanies each extra gluon in the
gluon ladder. 
We can simplify the integrand in \re{I}
for $z'=z$, 
to get 
\beq{z=z'}
{
I}\Big|_{z'\approx z}
= 4N_c \,\phi_\pi(z)\,\int\limits^{1}_{z}  
  \frac{dz'}{z'-z} F_\zeta(\zeta\frac{z'\bar z}{z'-z})
\simeq 4N_c \,\phi_\pi(z)\!\int\limits_\zeta^1
 \!\frac{dy}{y}\, F_\zeta(y,q^2_{\perp}).
\eeq
For a flat gluon distribution $F_\zeta(y) \sim {\rm const}$ at $y\to 0$,
and the integration gives $ {\rm const}\cdot \ln 1/\zeta $ which is the
above
mentioned logarithm.
The r.h.s. of (\ref{z=z'}) with the factor $2N_c/y$ appearing in \re{z=z'}
can be interpreted as the relevant limit of the DGLAP-type evolution equation
\cite{Rad96}
\begin{eqnarray}
\label{NNN}
&&q_\perp^2 \frac{\partial}{\partial q_\perp^2}\,F_\zeta(x=\zeta,q_\perp^2)
 = \nonumber \\ 
&&=\frac{\alpha_s}{2\pi} \int\limits_{\zeta}^1 dy\,P_\zeta^{gg}(\zeta,y)\,
F_\zeta(y,q_\perp^2)
 \simeq \frac{\alpha_s}{2\pi} \int\limits_{\zeta}^1 dy\,\frac{2N_c}{y}\,
F_\zeta(y,q_\perp^2)\,. 
\end{eqnarray}
The quantity on the l.h.s. of \re{NNN} defines what can be called the
unintegrated non-forward gluon distribution and the physical meaning of
Eqs.~\re{z=z'} and \ref{NNN} is that in the region $z'\sim z$ hard
gluon exchange can be viewed as a large transverse momentum part of the
gluon distribution in the proton, cf. \cite{NSS00}.

Next, consider the contribution to the imaginary part of the amplitude
of the dijet production coming from the end-points $z'\to 0$ and
$z'\to 1$: 
\beq{end}
 {
I}\Big|_{\rm end-points}
 =  \left(N_c+\frac{1}{N_c}\right)
 z \bar z \,  \int\limits^1_0 dz'\,
\frac{\phi_\pi(z',\mu^2)}{\bar z'^2}F_\zeta(\zeta,\mu^2) \,.
\eeq
Since $\phi_\pi(z')\sim z'$ at $z'\to 0$, the integral over $z'$
diverges logarithmically. This divergence indicates that the
collinear factorization conjectured in \re{factor} is generally not valid.
Remarkably, the divergent integral containing the pion distribution
amplitude
is just a constant and does not involve any $z$-dependence.
Therefore, the longitudinal momentum distribution of the jets in the
nonfactorizable contribution is calculable and, as it turns out,
has the shape of the asymptotic pion distribution amplitude
$\phi_\pi^{\rm as}(z) = 6z\bar z$.

Our result (\ref{I}) containing the end-point singularities 
 does not agree
with the result of independent calculations done in \cite{C} 
by means of the   method
which the light-cone dominance assumed from the very begin. We intend to
clarify the origin of this disagreement in a forthcoming publication.

\end{document}